\documentstyle[aps,pra,multicol,amssymb,amsfonts]{revtex}

\newcommand{\alg}[1]{\mbox{$\mathcal{#1}$}}

\begin{document}
\title{Complementarity between Position and Momentum \\ as a 
Consequence of 
Kochen-Specker Arguments}
\author{Rob Clifton}
\address{Departments of Philosophy and History and Philosophy of
Science, \\ 10th floor, Cathedral of
Learning, University of
Pittsburgh, \\ Pittsburgh, PA\ 15260, U.S.A. \\ email: rclifton+@pitt.edu}

\maketitle
\begin{abstract}  We give two simple Kochen-Specker arguments for 
complementary between the position and momentum components of spinless 
particles, arguments that are identical in structure 
to those given by Peres and Mermin for spin-1/2 particles. 
\end{abstract} \draft

\begin{multicols}{2}
\section{Introduction}

Complementarity is the idea that mutually exclusive pictures are 
needed for a complete description of quantum-mechanical reality.  The 
paradigm example is the complementarity 
between particle and wave (or `spacetime' and `causal') pictures, 
which Bohr took to be 
reflected in the uncertainty 
relation $\Delta x\Delta p\geq\hbar$.  Bohr saw this relation as defining the 
latitude of applicability of the concepts of position and momentum to 
a single system, not just as putting a limit on our ability to predict the values of both 
position and momentum within an ensemble of identically prepared systems.
Furthermore, right at the start of his celebrated reply to the 
Einstein-Podolsky-Rosen (EPR) argument 
against the completeness of quantum theory \cite{epr}, Bohr 
confidently asserted:
``it is never possible, in the description 
of the state of a mechanical system, to attach definite values to both 
of two canonically conjugate variables'' \cite{bohr}.
Critics have often pointed out that complementarity does not 
logically follow from the uncertainty relation without making the 
positivistic assumption that position and momentum can only be 
simultaneously defined if their values can be simultaneously measured 
or predicted 
\cite{popper}. 
However, we shall show here how direct Kochen-Specker arguments for complementarity between 
position and momentum can be given that are entirely independent of 
the uncertainty relation and its interpretation.

The aim of a Kochen-Specker argument is to establish that a certain 
set of observables of a quantum system cannot have
simultaneously definite values that respect the functional 
relations between compatible observables within the 
set \cite{ks}.  
Let $\alg{O}$ be a collection of bounded self-adjoint 
operators (acting on some Hilbert space) containing the identity $I$ and both $AB$ and $\lambda A+\mu B$  
($\lambda,\mu\in\mathbb{R}$), 
whenever $A,B\in\alg{O}$ and $[A,B]=0$.  Kochen 
and Specker called such a 
structure
a \emph{partial} algebra because there is no 
requirement that $\alg{O}$ contain 
\emph{arbitrary}  self-adjoint functions of its members (such 
as $i[A,B]$ or $A+B$, when $[A,B]\not=0$).  They then assumed that 
an assignment of values $[\cdot]:\alg{O}\rightarrow\mathbb{R}$ to the observables 
in $\alg{O}$ should at least be a \emph{partial} homomorphism,
respecting linear combinations and products of \emph{compatible} observables 
in $\alg{O}$.  That is, whenever $A,B\in\alg{O}$ and $[A,B]=0$, 
\begin{equation} \label{eq:constraints}
   \left[AB\right] = \left[A\right]\left[B\right],\ 
    \left[\lambda A+\mu B\right] = \lambda \left[A\right]+\mu 
    \left[B\right],\ [I]=1.
   \end{equation}
   Clearly these constraints 
   are motivated by analogy with classical physics, in which all 
   physical magnitudes (functions on phase space) trivially commute, 
   and possess values (determined by points of phase space) that respect 
   their functional relations.  (The requirement that $[I]=1$ is only needed to avoid triviality; for
    $[I^{2}]=[I]=[I]^{2}$ 
   implies $[I]=0$ or $1$, and if we took $[I]=0$, it would follow 
   that  
   $[A]=[AI]=[A][I]=0$ for all $\alg{A}\in\alg{O}$.)   
   
   Constraints (\ref{eq:constraints}) are not entirely out of place 
   in quantum theory.   For example, any common eigenstate $\Psi$ of a  collection of  observables $\alg{O}$ 
   automatically 
   defines a 
   partial homomorphism, given by assigning to each 
   $A\in\alg{O}$ the eigenvalue $A$ 
   has in state $\Psi$.  
   Difficulties --- called Kochen-Specker contradictions or 
   obstructions \cite{hamilton} --- arise when 
   not all observables in 
   $\alg{O}$ share a common eigenstate.  In that case, there is no 
   guarantee that 
   value assignments on all the commutative subalgebras of $\alg{O}$ 
   can be extended to a partial homomorphism on $\alg{O}$ as a whole.  Should such an 
   extension exist, one could be justified in thinking of the 
   noncommuting 
   observables in $\alg{O}$ as having simultaneously definite values, 
   notwithstanding that
   a quantum state may not permit all their values to be predicted 
   with certainty.  But should some 
   particular collection of observables $\alg{O}$ not possess 
   \emph{any} partial homomorphisms, the natural response would be to
   concede to Bohr that the observables 
   in $\alg{O}$ ``transcend the scope of classical physical 
   explanation'' and cannot be discussed using ``unambiguous language with suitable application of the 
   terminology of classical physics'' \cite{bohr2}.  
   That is, one would have strong reasons for 
   taking the noncommuting observables in $\alg{O}$ to be mutually 
   complementary.
   
  Bell 
   \cite{ks} has emphasized one other way to escape 
   an obstruction with respect to some set of 
   observables $\alg{O}$.  One could still take all $\alg{O}$'s observables to 
   have definite values by allowing the value of a particular 
   $A\in\alg{O}$ to be a function of the context in which $A$ is 
   measured.  Thus, suppose $\alg{O}_{1},\alg{O}_{2}\subseteq\alg{O}$ are 
   two different commuting subalgebras both containing $A$, where 
   $[\alg{O}_{1},\alg{O}_{2}]\not=0$.  Then if
   $[\cdot]_{1},[\cdot]_{2}$ are homomorphisms on these subalgebras 
   such that $[A]_{1}\not=[A]_{2}$, one could 
   interpret 
   this difference in values (the obstruction) 
   as signifying that the measured result for $A$ 
   has to depend on whether it is measured along with the observables in $\alg{O}_{1}$ 
   or those in $\alg{O}_{2}$.  
   Such 
   value assignments to the observables in $\alg{O}$ are called 
   contextual, because the context in which an observable is measured is allowed to 
   influence what outcome is obtained \cite{ks}.  For 
   example, Bohm's theory is 
   contextual in exactly this sense \cite{bohm}.  
   On the other hand (as Bell himself
   was quick to observe), complementarity also demands a kind of 
   contextualism: in some 
   contexts it is appropriate to assign a system a 
   definite position, and in 
   other contexts, a definite momentum.  The difference from
   Bohm is that Bohr takes the definiteness of the values of observables 
   \emph{itself} to be a function of context.  
   And this makes \emph{all} the difference 
   in cases where value contextualism can only be 
   enforced by making the measured value of an observable nonlocally depend on 
   whether an observable of another 
   spacelike-separated 
   system is measured.  We shall see below that complementarity between position 
   and momentum can only be avoided by embracing such nonlocality.  
      
   Numerous Kochen-Specker obstructions have been identified in the 
   literature, and their practical and theoretical 
   implications continue to be analyzed 
   \cite{cabello}.  
   While obstructions cannot
   occur for observables sharing a common eigenstate, failure to 
   possess a 
   common eigenstate does not suffice for an obstruction.  
   As Kochen and 
   Specker themselves showed, 
   the partial algebra generated by all components of a 
   spin-1/2 particle possesses plenty of partial homomorphisms.  
   But for particles with higher spin, or  
   collections of more than one spin-1/2 
   particle, obstructions can occur, perhaps the simplest being 
   those identified by Peres \cite{peres} in the case of two spin-1/2 particles, 
   and Mermin\cite{mermin} in the case of three.  Obstructions for sets that contain
   functions of position and momentum 
   observables \emph{have} been identified \cite{fleming}, but 
   additional observables need to be invoked that weaken the case for
    complementarity between position and momentum alone. 
    In the arguments below, we shall only need simple 
   \emph{continuous} functions of 
   the individual position and momentum components of a system.  
   Though all our observables have purely continuous spectra, obstructions 
   arise in 
   exactly the 
   same way that they do in the arguments given by Peres and Mermin 
   for the spin-1/2 case.  And because our 
   obstructions depend only on the structure of the Weyl 
   algebra, they immediately extend to relativistic
   quantum field theories, which are constructed out of representations 
   of the Weyl 
   algebra \cite{wald}.    
   
   \section{The Weyl Algebra}
   
 Let $\vec{x}=(x_{1},x_{2},x_{3})$ and 
 $\vec{p}=(p_{1},p_{2},p_{3})$ be the unbounded position and 
 momentum operators for 
 three degrees of freedom.   We cannot extract a Kochen-Specker 
 contradiction directly out of these operators, since 
 domain questions prevent them from defining a simple 
 algebraic structure.  However, we may just as well consider the collection of all 
 bounded, continuous, self-adjoint 
functions of $x_{1}$, and, similarly, the same set 
of functions in each of the variables $x_{2},x_{3},p_{1},p_{2},p_{3}$.  Taking $\alg{O}$ 
to be the partial algebra of observables generated by all these 
functions (obtained by taking compatible products and linear 
combinations thereof), we shall show that $\alg{O}$ does not possess any partial 
homomorphisms.  

Our arguments are greatly simplified by employing the following 
method, 
analogous to simplifying a problem in real analysis by passing to 
the complex plane.  Assuming that $\alg{O}$ \emph{does} possess a partial 
homomorphism $[\cdot]:\alg{O}\rightarrow\mathbb{R}$ (an assumption we shall eventually have to 
discharge), we can extend this mapping to the set 
$\alg{O}_{\mathbb{C}}\equiv\alg{O}+i\alg{O}$ in a well-defined manner, by taking
$[X]\equiv[\Re(X)]+i[\Im(X)]\in \mathbb{C}$, 
where $\Re(X)$ and $\Im(X)$ are the unique real and 
imaginary parts of $X$. Now, if we consider any pair of 
commuting unitary operators 
$U,U'\in\alg{O}_{\mathbb{C}}$, then since $U,U^{*},U',U'^{*}$ pairwise commute, 
the four self-adjoint operators
\begin{eqnarray}
\Re(U)=(U+U^{*})/2,\ \   & \Im(U)=i(U^{*}-U)/2, \\
 \Re(U')=(U'+U'^{*})/2,\ \  & 
\Im(U')=i(U'^{*}-U')/2,  
\end{eqnarray}
which must lie in $\alg{O}_{\mathbb{C}}$, also  pairwise commute.  Thus
\begin{eqnarray} \nonumber
UU' & = & \Re(U)\Re(U')-\Im(U)\Im(U') \\
& & +i(\Re(U)\Im(U')+\Im(U)\Re(U'))\in\alg{O}_{\mathbb{C}},
\end{eqnarray}
       using the fact that $\alg{O}$ is a partial algebra.  In 
       addition, 
\begin{eqnarray}
[UU'] & = &  
[\Re(U)\Re(U')-\Im(U)\Im(U')] \nonumber \\
& & +i[\Re(U)\Im(U')+\Im(U)\Re(U')], \\
& = & 
[\Re(U)][\Re(U')]-[\Im(U)][\Im(U')] \nonumber \\ \label{eq:next}
& & +i([\Re(U)][\Im(U')]+[\Im(U)][\Re(U')]), \\
& = & ([\Re(U)]+i[\Im(U)])([\Re(U')]+i[\Im(U')]) \\
& = & [U][U'],
\end{eqnarray}
using the fact that $[\cdot]$ is a partial 
homomorphism in step (\ref{eq:next}).  
So we have established that the following product rule must hold 
in $\alg{O}_{\mathbb{C}}$:
\begin{eqnarray}
& U,U'\in\alg{O}_{\mathbb{C}}\ \ \&\ \   & [U,U']=0 \nonumber \\
 \Rightarrow \   & UU'\in \alg{O}_{\mathbb{C}}\ \ \&\ \  & 
[UU']=[U][U'].
\end{eqnarray}
Henceforth, we shall only this need this simple product rule, together 
with $[\pm I]=\pm 1$.  Our 
obstructions will manifest themselves as contradictions obtained by 
applying the product rule to compatible unitary operators in 
$\alg{O}_{\mathbb{C}}$.  

To see what operators those are, we first 
recall the definition of the Weyl algebra for three degrees of 
freedom.   
Consider the two families of unitary operators 
given by
\begin{equation} \label{eq:Weyl}
U_{\vec{a}}=e^{-i\vec{a}\cdot\vec{x}/\hbar},\ 
V_{\vec{b}}=e^{-i\vec{b}\cdot\vec{p}/\hbar},\ 
\vec{a},\vec{b}\in \mbox{$\mathbb{R}$}^{3}.
\end{equation}
These operators act on any wavefunction $\Psi\in 
L_{2}(\mbox{$\mathbb{R}$}^{3})$ as 
 \begin{equation} \label{eq:one}
 (U_{\vec{a}}\Psi)(\vec{x}) = 
 e^{-i\vec{a}\cdot\vec{x}/\hbar}\Psi(\vec{x}),\   
  (V_{\vec{b}}\Psi)(\vec{x}) = \Psi(\vec{x}-\vec{b}), 
   \end{equation}
and satisfy the Weyl form of the canonical 
commutations relations $[x_{j},p_{k}]=\delta_{jk}i\hbar I$, 
\begin{equation} \label{eq:nocommute}
U_{\vec{a}}V_{\vec{b}} = e^{-i\vec{a}\cdot\vec{b}/\hbar}V_{\vec{b}}U_{\vec{a}}.
\end{equation}
The Weyl algebra (which is independent of the representation in 
(\ref{eq:one})) is just the $C^{*}$-algebra 
generated by the two families of unitary operators in 
(\ref{eq:Weyl}) subject to the commutation relation 
(\ref{eq:nocommute}).    

$\alg{O}_{\mathbb{C}}$ is properly contained in  the 
Weyl algebra.  Indeed, writing $U_{a_{j}}$ ($\equiv e^{-ia_{j}x_{j}/\hbar}$)
 for the $j$th component of the operator 
$U_{\vec{a}}$, and similarly $V_{b_{k}}$ ($\equiv e^{-ib_{k}p_{k}/\hbar}$), all nine 
of these 
component generators of the Weyl algebra lie in
$\alg{O}_{\mathbb{C}}$, because their real and imaginary parts, cosine 
and sine functions of the $x_{j}$'s and $p_{k}$'s, lie in $\alg{O}$.  
By the 
product rule, $\alg{O}_{\mathbb{C}}$ also contains the products of 
compatible unitary operators for different degrees of 
freedom, as well as compatible products of those products.  But, 
unlike the full Weyl algebra, $\alg{O}_{\mathbb{C}}$ 
does not contain incompatible products, like $U_{a_{j}}V_{b_{j}}$ when 
$a_{j}b_{j}\not=2n\pi\hbar$ ($n\in \mathbb{Z}$).  Nevertheless, $\alg{O}_{\mathbb{C}}$ is all we need to exhibit 
obstructions. The key is that we can choose values for the components 
of $\vec{a},\vec{b}$ so 
that, for $j=1$ to $3$, $a_{j}b_{j}=(2n+1)\pi\hbar$.  In that case, we immediately obtain 
from (\ref{eq:nocommute}) 
the \emph{anti}-commutation rule
\begin{equation} \label{eq:anti}
[U_{\pm a_{j}},V_{\pm b_{j}}]_{+}=0=[U_{\mp a_{j}},V_{\pm b_{j}}]_{+},
\end{equation}
which, together with the product rule, will generate the required 
obstructions. 

\section{Obstructions for Two and Three Degrees of Freedom}

We first limit ourselves to continuous functions of the four observables 
$x_{1},x_{2},p_{1},p_{2}$, extracting a contradiction in exactly the 
way Peres \cite{peres} does for a pair of spin-1/2 particles.  
 A first application of the product rule in $\alg{O}_{\mathbb{C}}$ 
 yields
 \begin{eqnarray} \label{eq:first'}
  \left[U_{-a_{1}}U_{a_{2}}\right] & = & \left[U_{-a_{1}}\right]\left[U_{a_{2}}\right], \\
 \left[U_{a_{1}}V_{b_{2}}\right] & = & \left[U_{a_{1}}\right]\left[V_{b_{2}}\right], \\
\left[V_{b_{1}}U_{-a_{2}}\right] & = & \left[V_{b_{1}}\right]\left[U_{-a_{2}}\right], \\
\left[V_{-b_{1}}V_{-b_{2}}\right] & = & \left[V_{-b_{1}}\right]\left[V_{-b_{2}}\right]. \label{eq:last'}
\end{eqnarray}
Multiplying equations (\ref{eq:first'})--(\ref{eq:last'}) together, and using one 
further (trivial) application of the product rule
\begin{equation} \label{eq:trivial}
 [U_{a_{j}}][U_{-a_{j}}]=[I]=1=[V_{b_{k}}][V_{-b_{k}}],
 \end{equation}
  one obtains
\begin{equation} \label{eq:square'}
\left[U_{-a_{1}}U_{a_{2}}\right]\left[V_{-b_{1}}V_{-b_{2}}\right]
\left[U_{a_{1}}V_{b_{2}}\right]\left[V_{b_{1}}U_{-a_{2}}\right] = 1.
 \end{equation}
 However, because of the anti-commutation rule (\ref{eq:anti}), the first pair of product 
 operators occurring in (\ref{eq:square'}) actually \emph{commute}, as do the second 
 pair of product operators.  
 Hence we may make a further application of the 
 product rule to (\ref{eq:square'}) to get 
\begin{equation} \label{eq:square''}
    \left[U_{-a_{1}}U_{a_{2}}V_{-b_{1}}V_{-b_{2}}\right]
    \left[U_{a_{1}}V_{b_{2}}V_{b_{1}}U_{-a_{2}}\right] = 1.
 \end{equation}
 Again, due to the anti-commutation rule, the  two remaining (four-fold) product operators occurring 
 in (\ref{eq:square''}) commute, and their product is  
 $-I$.  Thus, a final application of the 
 product rule to (\ref{eq:square''}) yields the contradiction 
 $[-I]=-1=1$.
 
 Notice that this obstruction remains for 
 any given nonzero values for $a_{1}$ and $a_{2}$, provided only that we choose 
 $b_{1,2}=(2n+1)\pi\hbar/a_{1,2}$.  The obstruction would vanish if, 
 instead, we chose any of the numbers $a_{1},a_{2},b_{1},b_{2}$ to be zero.   
 When $a_{1}=a_{2}=0$ or  $b_{1}=b_{2}=0$, this is to be expected, since 
 one would then no longer be attempting to assign values to nontrivial functions 
 of \emph{both} the positions and momenta.  However, the breakdown of the 
 argument when either $a_{2}$ or $b_{2}$ is zero does not necessarily 
 mean that a more complicated argument could not be given 
 for position-momentum complementarity by invoking 
  only a \emph{single} degree of freedom. 
 
 As Mermin \cite{mermin} has emphasized (for the spin-1/2 analogue of 
 the above argument), one can get by without 
 independently assuming the existence of values for the two commuting unitary operators 
 occurring in (\ref{eq:square''}), and 
 thereby strengthen the argument.  For we can suppose that the quantum state 
 of the system is an eigenstate of these operators, with 
 eigenvalues that \emph{necessarily} multiply to -1. 
 Using (\ref{eq:one}), a wavefunction $\Psi$ will 
 be an eigenstate of both products in (\ref{eq:square''}) just in case 
 \begin{eqnarray} \label{eq:hi}
 e^{i(a_{1}x_{1}-a_{2}x_{2})/\hbar}\Psi(x_{1}+b_{1},x_{2}+b_{2})=c\Psi(x_{1},x_{2}), \\
 -e^{-i(a_{1}x_{1}-a_{2}x_{2})/\hbar}\Psi(x_{1}-b_{1},x_{2}-b_{2})=c'\Psi(x_{1},x_{2}),  
 \label{eq:hi'}
 \end{eqnarray}
 for some $c,c'\in\mathbb{C}$.  We should not 
 expect there to be a \emph{normalizable} wavefunction satisfying (\ref{eq:hi}) and (\ref{eq:hi'}),
 because the commuting product operators in (\ref{eq:square''}) have purely continuous spectra.  
 But if we allow ourselves the idealization of using Dirac 
 states (which can be approximated arbitrarily closely by 
 elements of $L_{2}(\mbox{$\mathbb{R}$}^{2})$), and just choose 
 $a_{1}=a_{2}$ for simplicity, then the two-dimensional delta function 
 $\delta(x_{1}-x_{2}-x_{0})$ --- an improper eigenstate of the 
 relative position operator $x_{1}-x_{2}$ with `eigenvalue' 
 $x_{0}\in\mathbb{R}$ --- provides a simple solution to the above 
 equations.  However, this
 state cannot also be used to independently justify the assignment of values 
 to the operators $U_{a_{1}}V_{b_{2}}$ and $V_{b_{1}}U_{-a_{2}}$
  occurring in (\ref{eq:square'}), which do not have 
  $\delta(x_{1}-x_{2}-x_{0})$ as an eigenstate. 
 
 It is ironic that $\delta(x_{1}-x_{2}-x_{0})=\delta(p_{1}+p_{2})$ is exactly 
 the state of two spacelike-seperated particles that EPR invoked to argue \emph{against} 
 position-momentum complementarity.  So in a sense the EPR argument 
 carries the seeds of its own destruction.  For  suppose we follow their 
 reasoning by invoking locality and the strict correlations entailed 
 by the EPR state between $x_{1}$ and $x_{2}$, and between $p_{1}$ and 
 $p_{2}$,  
 to argue for the existence of noncontextual 
 values for all four positions and momenta.  Then all eight component unitary 
 operators we employed above must have definite noncontextual values, since 
 their real and imaginary parts are simple functions of those 
 $x$'s and $p$'s.  It 
 is then a small step to conclude that the four product operators in (\ref{eq:square'}) 
 should also have
 definite noncontextual values satisfying the product rule, and from there 
 contradiction follows.  This final step cannot itself be 
 justified by appeal to locality, for the four product observables 
 in (\ref{eq:square'}) do not 
 pertain to either particle on its own and, hence, a measurement 
 context for any one of these operators (i.e., their self-adjoint real and 
 imaginary parts) necessarily requires a joint measurement undertaken on both 
 particles \cite{mermin}.  Still, the above argument sheds an entirely new 
 light on the nonclassical features of the original EPR 
 state, which have hitherto only been discussed from a statistical 
 point of view \cite{bell}.
  
  Our second argument employs all three degrees of freedom, extracting a contradiction in exactly the 
way Mermin \cite{mermin} does for three spin-1/2 particles.
  Again, a first application of the product rule in $\alg{O}_{\mathbb{C}}$ 
 yields
 \begin{eqnarray} 
\left[U_{a_{1}}V_{-b_{2}}V_{-b_{3}}\right]  
& 
= & \left[U_{a_{1}}\right]\left[V_{-b_{2}}\right]\left[V_{-b_{3}}\right], \label{eq:first} \\ 	
\left[V_{-b_{1}}U_{a_{2}}V_{b_{3}}\right] & = &  
\left[V_{-b_{1}}\right]\left[U_{a_{2}}\right]\left[V_{b_{3}}\right], \\
\left[V_{b_{1}}V_{b_{2}}U_{a_{3}}\right] 
& = &  
\left[V_{b_{1}}\right]\left[V_{b_{2}}\right]\left[U_{a_{3}}\right], \\
 \left[U_{-a_{1}}U_{-a_{2}}U_{-a_{3}}\right] 
 & = & \left[U_{-a_{1}}\right]\left[U_{-a_{2}}\right]
 \left[U_{-a_{3}}\right]. \label{eq:last}
 \end{eqnarray} 
 Multiplying (\ref{eq:first})--(\ref{eq:last}) together, again 
 using (\ref{eq:trivial}), 
  yields
 \begin{eqnarray} \nonumber
    \left[U_{a_{1}}V_{-b_{2}}V_{-b_{3}}\right]\left[V_{-b_{1}}U_{a_{2}}V_{b_{3}}\right] & & \\ \label{eq:square}
   \left[V_{b_{1}}V_{b_{2}}U_{a_{3}}\right]\left[U_{-a_{1}}U_{-a_{2}}U_{-a_{3}}\right] & = & 1.
 \end{eqnarray}
 But now, exploiting the anti-commutation rule once again, the four 
 product operators occurring
 in square brackets in (\ref{eq:square}) pairwise commute, and 
 their product is easily seen to be 
 $-I$.  
 So one final application of the 
 product rule to (\ref{eq:square}) once more yields the 
 contradiction $[-I]=-1=1$.
  
As before, we may interpret the $x$'s and $p$'s as the positions 
 and momenta of three spacelike-separated particles.  And we can
  avoid independently assuming values for the four 
 products in (\ref{eq:square}) by taking the state of the 
 particles to be a simultaneous (improper) eigenstate of these 
 operators --- exploiting that state's strict correlations and 
 EPR-type reasoning from locality to motivate values for all 
 the component operators. (The reader is invited to use (\ref{eq:one}) 
 to determine the set of all such common 
 eigenstates, which are new position-momentum analogues of the 
 Greenberger-Horne-Zeilinger state \cite{mermin}.)     
 This time, the \emph{only} way to prevent contradiction is to 
 introduce contextualism to distinguish, for example, the 
 value of $U_{a_{1}}$ as it occurs in (\ref{eq:first}) from the 
 value this operator (or rather its inverse) receives in 
 (\ref{eq:last}) in the context of different 
 operators for particles $2$ and $3$ --- forcing the values 
 of $\sin a_{1}x_{1}$ and $\cos a_{1}x_{1}$ to  
 \emph{nonlocally} depend on whether position or momentum observables for 
 $1$ and $2$ are measured.
Bohr of course denied
  that there could be any such nonlocal ``mechanical'' 
  influence, but only ``an influence on the very conditions which 
  define'' which of the two mutually complementary pictures available 
  for each system can be unambiguously employed \cite{bohr}.    
 
The author would like to thank All Souls College, Oxford for support, and Paul Busch, Jeremy Butterfield, and 
Hans Halvorson for helpful 
discussions.

\end{multicols}
\end{document}